\documentclass[aps,amsmath,amssymb,twocolumn,superscriptaddress,preprintnumbers,nofootinbib]{revtex4-1}

\usepackage{graphicx}
\usepackage{dcolumn}
\usepackage{bm}
\usepackage{hyperref}
\usepackage{epstopdf}
\usepackage{color}
\newcommand{\nc}{\newcommand}

\newcommand{\met}{E\!\!\!/_T}
\newcommand{\Eq}[1]{Eq.~(\ref{#1})}
\newcommand{\Sec}[1]{Sec.~\ref{#1}}
\newcommand{\App}[1]{App.~\ref{#1}}
\newcommand{\Fig}[1]{Fig.~\ref{#1}}
\newcommand{\Tab}[1]{Table~\ref{#1}}
\newcommand{\Ref}[1]{Ref.~\cite{#1}}

\def\lsim{\raisebox{-4pt}{$\,\stackrel{\textstyle{<}}{\sim}\,$}}
\def\gsim{\raisebox{-4pt}{$\,\stackrel{\textstyle{>}}{\sim}\,$}}
\newcommand{\be}{\begin{equation}}
\newcommand{\ee}{\end{equation}}
\newcommand{\bi}{\begin{itemize}}
\newcommand{\ei}{\end{itemize}}
\def\beq{\begin{equation}} 
\def\eeq{\end{equation}} 
\def\bea{\begin{eqnarray}} 
\def\eea{\end{eqnarray}} 
\def\ben{\begin{enumerate}} 
\def\een{\end{enumerate}}

\def\lsim{\mathrel{\raise.3ex\hbox{$<$\kern-.75em\lower1ex\hbox{$\sim$}}}} 
\def\gsim{\mathrel{\raise.3ex\hbox{$>$\kern-.75em\lower1ex\hbox{$\sim$}}}} 
\def\ifmath#1{\relax\ifmmode #1\else $#1$\fi}

\nc{\mc}{\mathcal}
\nc{\ttbar}{t\bar t}
\newcommand{\tafb}{{\cal A}^{t \bar{t}}}

\begin{document}
\DeclareGraphicsExtensions{.jpg,.pdf,.mps,.png,}

\title{A Polarized View of the Top Asymmetry}


\author{David Krohn}
\email{dkrohn@physics.harvard.edu}
\affiliation{Department of Physics, Harvard University, Cambridge MA, 02138}

\author{Tao Liu}
\email{taoliu@physics.ucsb.edu}
\affiliation{Department of Physics, University of California,
Santa Barbara CA, 93106}

\author{Jessie Shelton}
\email{j.shelton@yale.edu}
\affiliation{Department of Physics, Yale University, New Haven CT, 06511}

\author{Lian-Tao~Wang}
\email{liantaow@uchicago.edu}
\affiliation{Department of Physics, Enrico Fermi Institute, and Kavli Institute for Cosmological Physics, \\ University of Chicago, Chicago IL, 60637 }

\begin{abstract}
Recent experimental results from the CDF collaboration which study the top forward-backward asymmetry have strengthened the case that new physics is playing a role in $t\bar t$ production.  Here we propose a set of measurements, built from the charged lepton kinematics in semileptonic and fully leptonic $\ttbar$ events, designed to further probe the underlying causes of this asymmetry both at the Tevatron and at the LHC.  Using a set of conservative reference models we find that measurements of the charged lepton asymmetry, top polarization, and $t\bar t$ spin correlation can establish the existence of new physics and distinguish between competing models both at the Tevatron and the LHC.  At the Tevatron, discrimination between models is possible at the $3\sigma$ level.  At the LHC, we demonstrate that a top forward-backward asymmetry can be established at $\gsim 3\sigma$ in the first $\sim5\ {\rm fb}^{-1}$ of data, and show how competing explanations can be further disentangled.

%

\end{abstract}

\maketitle

\section{Introduction}
\label{sec:intro}
Roughly three years ago both CDF and D0 reported anomalously large top
forward-backward asymmetries (AFB)~\cite{Aaltonen:2008hc,*DO:2007qb}.
Since that time analyses have improved and integrated luminosities
have increased such that the latest semileptonic
results~\cite{Aaltonen:2011kc} from CDF report a $3.4 \sigma$
discrepancy with the Standard Model (SM)
prediction~\cite{Bowen:2005ap,Kuhn:1998kw,*Antunano:2007da,*Almeida:2008ug}
in the high $m_{t\bar{t}}$ region, while in the fully leptonic\footnote{Here and throughout this 
paper we will use ``leptonic'' to refer only to electrons and muons.} channel
recent measurements~\cite{CDF-leptonic:2011} display a $2.6
\sigma$  deviation in the same direction.

During this time the theory community has dedicated much effort to
model building aimed at explaining these
excesses~\cite{Djouadi:2009nb,*Martynov:2009en,*Ferrario:2009bz,*Jung:2009jz,*Cheung:2009ch,*Frampton:2009rk,*Shu:2009xf,*Arhrib:2009hu,*Dorsner:2009mq,*Barger:2010mw,*Cao:2009uz,*Cao:2010zb,*Xiao:2010hm,*Martynov:2010ed,*Chivukula:2010fk,*Rodrigo:2010gm,*Bauer:2010iq,*Zhang:2010dr,*Chen:2010hm,*Xiao:2010ph,*Alvarez:2010js,*Burdman:2010gr,*Cheung:2011qa,*Bai:2011ed,*Shelton:2011hq,*Barger:2011ih,*Blum:2011up,*Grinstein:2011yv,*Patel:2011eh,*Zerwekh:2011wf,*Delaunay:2011gv,*Ligeti:2011vt,*Kagan:2011yx,*Jung:2011zv,*Buckley:2011vc,*Shu:2011au,*Rajaraman:2011rw,
*AguilarSaavedra:2011zy,*Chen:2011mg,*Nelson:2011us,*Wang:2011ta,*Anchordoqui:2011ag,
*Jung:2011ua,*Zhu:2011ww,*Fox:2011qd,*Jung:2011ue,*Babu:2011yw,*Cao:2011yt,*Djouadi:2011aj}. The
good agreement of the top pair production cross-section with SM
predictions drives most models to generate the asymmetry from
interference between new physics and the standard model, leading to
the introduction of either a flavor-nonuniversal $s$-channel colored
vector boson or a $t$-channel color singlet with nontrivial flavor
structure.

If physics beyond the Standard Model (BSM) is really responsible for
the top asymmetry then it would be desirable to probe its particle content
directly by producing any new states on-shell and then measuring their
masses and couplings.  Depending on the model, such searches may proceed
in $\ttbar$ resonances or top-light jet resonances, and interesting signals may be discovered
shortly.  However, excesses in the {\em nonresonant} $\ttbar$ cross-section
alone allow for observations of departures from the SM~\cite{Degrande:2010kt,*AguilarSaavedra:2011vw}.  Here we establish that 
a BSM leptonic charge asymmetry can also be seen in nonresonant $\ttbar$
events at the 7 TeV LHC.  By virtue of the fact that the BSM states must have couplings strong
enough to impact the measurement of AFB in the low to intermediate
range of $m_{t\bar t}$, we can confidently measure some essential
properties of relevant BSM states in a fairly model-independent way.
Chief among these is the chiral nature of their couplings to the top,
and the correlation between top polarizations in the final
state\footnote{For recent proposals focusing on other observables, see
Refs~\cite{Wang:2010du,*Xiao:2011kp,*Cao:2011ew,*Bhattacherjee:2011nr,Craig:2011an,Gresham:2011pa,Hewett:2011wz,Jung:2009pi}.}.
We further study the asymmetry with leptonic variables;
together with measurements of spin correlations and polarization, models
for the top asymmetry can be well tested within the 7 TeV run.  

Measuring the chiral nature of the top's coupling in the region where
we measure AFB would be useful  because this coupling is
vector-like at tree level in the SM, so any deviation from this, as is
required in e.g. models with an $s$-channel color vector octet, would
indicate the presence of new physics.  Moreover, the strong constraints on
$b$ quark couplings from precision flavor observables leads many models 
for AFB to couple only to right-handed tops, predicting large and distinct
polarization signals.
Any confirmation of a non-vector-like coupling, especially one which
is dominantly right-handed, would strengthen the case for new physics.
Furthermore, the correlation between the top and anti-top
polarizations can be used to distinguish between different production
mechanisms, particularly at the LHC, where BSM contributions to the
top pair production cross-section are expected to be large.

Fortunately, as the top decays before hadronization effects can
wash out its polarization information, many useful spin observables can be constructed in 
$\ttbar $ events.  It has been known for more
than two decades that the differential distributions of top decay
products serve as an excellent probe of new
physics~\cite{Barger:1988jj,*Kane:1991bg,*Schmidt:1992et,*Brandenburg:1992be,*Jezabek:1994zv,*Mahlon:1995zn},
with many applications to the study of $t\bar t$
resonances~\cite{Agashe:2006hk,*Lillie:2007yh,*Krohn:2009wm,*Baumgart:2011wk,*Berger:2011hn,*Djouadi:2007eg},
and, more recently, to distinguishing potential explanations of the
anomalous top AFB~\cite{Godbole:2010kr,*Jung:2010yn,*Choudhury:2010cd,*Cao:2010nw}.

The main contribution of this article is the study of the power of
various leptonic observables for determining the existence of an
enhanced top forward-backward asymmetry and discriminating amongst
competing explanations for it, using a set of relevant models and (for
the LHC) a realistic top-reconstruction algorithm.  We will employ a
set of four phenomenological reference models to inject signals
characteristic of the two basic classes of models constructed to
explain the forward-backward asymmetry: (1) models with an $s$-channel
vector boson are represented by three reference axigluon models with
fully axial, left-handed, and right-handed couplings, while (2)
$t$-channel vector boson models are represented by a reference $W'$.
Full details of our benchmarks can be found in \App{sec:models}, but
it is important to note that the particular models at hand (i.e. the
chosen masses and couplings) are not essential to our conclusions -
other $s$- and $t$-channel models constructed to yield a similar
asymmetry would yield lead us to similar conclusions.

We first consider the prospects at the Tevatron in section
\Sec{sec:tevatron}, examining the potential impact of leptonic
observables with current and projected luminosities at the Tevatron.
We emphasize that leptonic observables access novel information beyond
that contained in the distributions of the parent tops, and compare
the utility of measuring polarization directly versus indirectly,
through lepton charge asymmetries
\cite{Bowen:2005ap,*Melnikov:2009dn,*Bernreuther:2010ny}.  In
\Sec{sec:lhc} we will build upon prior
works~\cite{Godbole:2010kr,*Choudhury:2010cd,*Cao:2010nw} and present
a detailed set of cuts and observables which allow for useful probes
of relevant BSM physics within the first $1-5\ {\rm fb}^{-1}$ of
$7~{\rm TeV}$ LHC data.  In particular, we will see that a leptonic
charge asymmetry in dileptonic $\ttbar$ events can be established at
$\geq 3 \sigma$ in $5 {\rm fb}^{-1}$ for all of our BSM reference
models.  \Sec{sec:conclusions} contains our conclusions.  Finally, in 
\App{sec:reco} we give an
overview of our top reconstruction procedure,
\App{sec:models} discusses our benchmark models, and \App{sec:tables} 
contains many tabulated results.

\section{Tevatron Analysis}
\label{sec:tevatron}
As was mentioned in \Sec{sec:intro}, the polarization of the top is
reflected in the kinematic distributions of its daughters, as the top
decays before hadronization effects can wash away this information.
Thus top polarization $\mc{P}_{n}$ along a chosen axis $\hat n$ can be
measured by the angular distribution of the top decay products with
respect to that axis, measured in the top rest
frame~\cite{Brandenburg:2002xr,*Shelton:2008nq}:
\be
\label{eq:spinbasic}
\frac{1}{\Gamma}\frac{d\Gamma}{d\cos\theta_{i,n}}=
        \frac{1}{2}\left(1+\mc{P}_{n}\kappa_i\cos\theta_{i,n}\right)
\ee
where $\mc{P}_{n}=\pm 1$ for tops completely polarized (anti-) parallel
with the chosen axis, $\kappa_i$ is the {\it spin analyzing power} of
decay product $i$, and $\theta_i$ is the direction of each
daughter with respect to the chosen axis, as measured in the rest
frame of the top.  For the $b$ we have $\kappa=-0.4$, while for the
neutrino $\kappa=-0.3$, and the charged lepton has $\kappa=1.0$.
Thus, of all the particles coming from the decay of the top the
charged lepton is most sensitive to the top's polarization.

The high sensitivity of the charged lepton is convenient, because of
all the top decay products it is the easiest to identify and measure.
The purpose of this section is to point out that simple variables
constructed from the leptons in semileptonic and dileptonic $\ttbar $
events have hitherto untapped power to distinguish between competing
explanations of the observed asymmetry even at the Tevatron, and have
the potential to significantly strengthen the case for new physics
beyond the standard model.
Models which attempt to explain the observed top asymmetry typically
predict heavy new states with nontrivial chiral structure.  This
translates into a potentially large net polarization of tops as well
as a departure from SM spin correlations, and therefore potentially
large signals in the lepton distributions which are capable of
distinguishing between different models \cite{Choudhury:2010cd}.

For a given choice of axis $\hat n$, the net polarization of the top
sample along that axis can be extracted as
\beq
\label{eq:pol}
\mc{P}_{n} = \frac{N(\cos\theta_{\ell,n} > 0) - N(\cos\theta_{\ell,n}
  < 0) } {N(\cos\theta_{\ell,n} > 0) + N(\cos\theta_{\ell,n} < 0)}.
\eeq Three commonly considered polarization bases are (1) $\mc{P}_{\rm
  h}$, the helicity basis, where $\hat n$ is given by the direction of
the parent top's momentum in the CM frame; (2) $\mc{P}_{\rm b}$, the
beam basis, where $\hat n$ is given by the direction of the beam; and
(3) $\mc{P}_{\rm off-d}$, the off-diagonal axis \cite{Parke:1996pr},
which maximizes SM spin correlations and interpolates between the
previous two.  The SM predicts a small net polarization arising from
electroweak corrections to top pair production, which we
neglect\footnote{The corrections obtained from SM EW processes will at
  most shift our observables by a small linear amount and will not
  have any qualitative effect on our conclusions.}.  After imposing
selection cuts~\cite{Aaltonen:2011kc,CDF-leptonic:2011} as discussed
below, however, SM tops will in general show a nonzero polarization.
In \Tab{tab:tevpol1} we display these values as well as predictions
for our reference models in the helicity basis.  Results for the beam
and off-diagonal bases are shown in \Tab{tab:tevpol2} and
\Tab{tab:tevpol3} in the Appendix.  The helicity basis gives better
separation between BSM models and the SM at the Tevatron than either
the beam basis or the off-diagonal basis.  This is not surprising: the
helicity basis becomes optimal when the top mass is small compared to
its energy, while the beam basis is effective when the top is
traveling with a small velocity, precisely where contributions from
BSM physics are smallest relative to the standard model.  Top
polarization from new physics will be larger at higher invariant mass
where the helicity basis is better suited. The off-diagonal basis,
which interpolates between the beam basis and the helicity basis, is
intermediate in sensitivity.

\begin{table}
\begin{center}
\begin{ruledtabular}
\begin{tabular}{c|cc|cc}
       & \multicolumn{2}{c|}{Semileptonic}   & \multicolumn{2}{c}{Dileptonic} \\
       & sel. cuts & $m_{t\bar t}>450~{\rm GeV}$ & sel. cuts & $m_{t\bar t}>450~{\rm GeV}$ \\
\hline
SM    & 4 \% (3 \%) & 7 \% (5 \%) & 4 \% (6.5 \%) & 6 \% (10 \%) \\     
$G_A$ & 5 \%  &  7 \% & 5 \% & 7 \% \\
$G_L$ & 2 \%  & -1 \% & 1 \% & -1 \%  \\
$G_R$ & 8 \%  & 12 \% & 8 \% & 12 \% \\
$W'$  & 15 \%  & 22 \% &  14 \% & 21 \% \\
\end{tabular}
\end{ruledtabular}
\end{center}
\caption{Net polarization ${\cal P}_ {\rm h}$ in the 
helicity basis at the Tevatron.   We note that in the SM, at tree-level, these asymmetries are all zero.  In parentheses are $1 \sigma$ statistical
errors uncertainties on an asymmetry measurement centered about the predicted 
SM value assuming $5.3 \mathrm{fb}^{-1}$ (semileptonic) or 
$5.1 \mathrm{fb}^{-1}$ (dileptonic).  Note that the effects of the differing semileptonic and dileptonic selection cuts
are small.
\label{tab:tevpol1}}
\end{table}

\begin{figure}[t]
\includegraphics[scale=1.05]{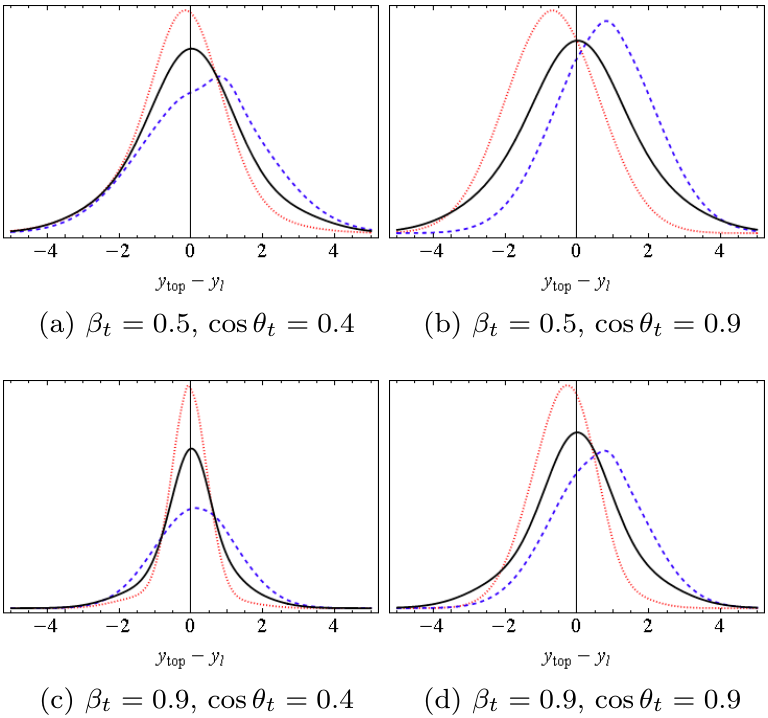}
\caption{ Distributions of the rapidity difference $y_{top}-y_{\ell}$ for right-handed 
(red, dotted), left-handed (blue, dashed)
and unpolarized (black, solid) tops, for fixed top kinematics as determined by the 
top boost $\beta_t $ and CM frame production angle $\cos\theta_t$.
\label{fig:top-lepton}}
\end{figure}

The lepton polarization angle $\cos\theta_\ell$ has the nice
feature that it is completely uncorrelated with the kinematics of the
parent tops as it is measured in the top rest frame.  However, reconstructing this frame 
is non-trivial and
can be difficult. It is possible to define other variables which use the
same underlying information, but may prove more flexible.  One especially
interesting variable is the {\it leptonic charge asymmetry} 
\cite{Bowen:2005ap,*Melnikov:2009dn,*Bernreuther:2010ny}
\beq
\label{eq:AEll}
{\cal A}^\ell_{FB} = \frac{N(q_\ell y_\ell > 0) - N(q_\ell y_\ell < 0)}
                           {N(q_\ell y_\ell > 0) + N(q_\ell y_\ell < 0)}
\eeq
in semileptonic events.  The charged lepton rapidity (in either the
lab or the CM frame) depends on the velocity $\beta_t$ and CM frame
production angle $\cos\theta_t$ of the semileptonic top as well as on
$\cos\theta_\ell$, but is independent of the lepton energy in the
top rest frame (as the lepton is effectively massless, and so the energy only 
changes the magnitude of its four-vector).  Thus the lepton asymmetry of Eq.~(\ref{eq:AEll}) is
an alternate measure of the lepton polarization: it contains
additional information about the top production mechanism, beyond the
information in the top AFB.  We illustrate the relationship between
top and lepton rapidities in \Fig{fig:top-lepton}.  For
dileptonic tops, one can define the {\it dileptonic} charge asymmetry,
\beq
\label{eq:ADeltaEll}
{\cal A}^{\Delta\ell}_{FB} = \frac{N( (y_{\ell^ +}- y_{\ell ^-}) > 0) - N( (y_{\ell^ +}- y_{\ell ^-}) <0 )}
                                  {N( (y_{\ell^ +}- y_{\ell ^-}) > 0) + N( (y_{\ell^ +}- y_{\ell ^-}) < 0)}
\eeq
which is frame-independent.
%
%
As can be seen in \Fig{fig:top-lepton}, the dependence of
lepton rapidity on parent top polarization is enhanced in the forward
regions.  Therefore a new source of right-handed tops which
preferentially populates high rapidity regions will lead to a
significant enhancement of forward leptons.  As central lepton
acceptance at the Tevatron extends only to $|\eta | <1.1$, this can
lead to marked acceptance differences between BSM and SM tops, as well
as differences between the BSM models themselves.  In particular,
acceptances need to be understood separately for $t$-channel
right-handed models and $s$-channel axially-coupled models (see also
\cite{Gresham:2011pa}).

While the correlation between the lepton asymmetries ${\cal
A}^\ell_{FB},\,{\cal A}^{\Delta\ell}_{FB} $ and the top
forward-backward asymmetry ${\cal A}^t_{FB}$ is less direct than a
direct measurement of the top polarization through the top rest frame
variable of \Eq{eq:pol}, the lepton asymmetry does not require
reconstruction of rest frames and can therefore provide a cleaner
probe of the system.  Moreover, as the lepton is not colored, a
measurement of the lepton asymmetry allows for easier comparison to
predictions from NLO QCD\footnote{Thanks to K.~Melnikov for
emphasizing this point.}.  In the absence of any polarization, a given
top asymmetry ${\cal A}^t_{FB}$ will produce a mildly reduced ${\cal
A}^\ell_{FB}$.  Left-handed tops, however, will yield a much reduced
${\cal A}^\ell_{FB}$, while for right-handed tops the leptonic
asymmetry is comparable to or greater than the parent top
asymmetry. If the tops are sufficiently forward that finite lepton
acceptance becomes relevant (i.e., if the top is sufficiently forward
then events which would contribute to the asymmetry will not pass
selection cuts - we will see this for the $W'$ model) then
this conclusion will not hold.

In the following subsections we consider semileptonic and dileptonic tops at the Tevatron
in more detail, focusing on lepton charge asymmetries.  All results are simulated using the full $2\rightarrow 6$ 
matrix elements of $\ttbar$ production as computed in 
Madgraph~\cite{Alwall:2007st}.

\subsection{Semileptonic Tops}

For semileptonic tops, we impose selection cuts after
\cite{Aaltonen:2011kc}, requiring: a charged lepton with $|\eta |
<1.0$ and $p_{T\ell} >20$ GeV; at least four jets with $p_{Tj} >40$
GeV and $|\eta | <2.0$; and missing energy $\met >20$ GeV.  Further,
at least one jet must be $b$-tagged, which means the tagged jet must
be central, $|\eta_b | <1.0$.  After these cuts, non-top background is
less than ${\cal O} ( 20\%)$ of remaining events; we neglect it here.
In what follows we will present results with statistical 
errors derived from the published number of events measured at CDF.

\begin{table}
\begin{center}
\resizebox{90mm}{!}{
\begin{tabular}{cl|c|c|c}
\hline \hline
       & frame and &$\ttbar$   & Lepton    & stat. sig. \\
       & mass range & asymmetry  & asymmetry & (5.3 fb$^{-1}$) \\
\hline
%
%
$G_A$         & lab, sel. cuts & 9 \% & 4 \% & 1.1    \\
              & lab, $m_{t\bar t}>450~{\rm GeV}$ & 17 \%  & 9 \% & 1.9 \\
              & $CM$, sel. cuts & 12 \% & 6 \% & 1.7 \\
              & $CM$, $m_{t\bar t}>450~{\rm GeV}$ & 19 \% & 12 \% & 2.4 \\

\hline

$G_L$         & lab, sel. cuts & 7 \%  &  -3 \%  & 0.9     \\
              & lab, $m_{t\bar t}>450~{\rm GeV}$ &  14 \%  &  -1 \% & 0.2\\
              & $CM$, sel. cuts & 13 \%   & -4 \%  & 1.4 \\
              & $CM$, $m_{t\bar t}>450~{\rm GeV}$ & 20\%  & -3 \%   & 0.6    \\

\hline

$G_R$         & lab, sel. cuts & 9 \%  & 12 \%  & 3.9  \\
              & lab, $m_{t\bar t}>450~{\rm GeV}$ &  14 \%  & 18 \% & 5.0 \\
              & $CM$, sel. cuts & 9 \%  & 16  \% & 3.5 \\
              & $CM$, $m_{t\bar t}>450~{\rm GeV}$ & 15 \%  & 22  \% & 4.4 \\

\hline

$W'$ & lab, sel. cuts & 15 \% & 13 \%  & 3.9    \\
     & lab, $m_{t\bar t}>450~{\rm GeV}$ & 26 \% & 22 \% & 4.9      \\
     & $CM$, sel. cuts & 20 \%  & 16 \%  & 4.4    \\
     & $CM$, $m_{t\bar t}>450~{\rm GeV}$ &  31 \% & 26 \%  & 5.3  \\

 \hline\hline
\end{tabular}
}
\end{center}
\caption{BSM contributions to the parton level $\ttbar $ and 
leptonic asymmetries after imposing CDF semileptonic acceptance cuts. 
Lepton asymmetries computed using both the lab and $CM$ frame lepton
rapidities are shown.  We note that in the SM, at tree-level, these asymmetries are all zero.
Statistical significances of the leptonic asymmetries are
based on the number of events observed in \cite{Aaltonen:2011kc}. 
\label{tab:tevsemilepeta} 
}
\end{table}

In \Tab{tab:tevsemilepeta}, we compare the leptonic charge asymmetry
to the parent top asymmetry in our reference models. The contribution
to the asymmetries from LO SM $\ttbar $ is zero for all asymmetries
even after selection cuts.  We quote statistical significance of the
leptonic asymmetries based on the number of events observed in 5.3
$\mathrm{fb}^{-1}$.  By the end of the Tevatron's run, twice this
amount of data will be available (for each experiment); as the
statistical uncertainties scale approximately as $1/\sqrt{N} $, the
maximum statistical reach of the Tevatron is larger by nearly a factor
of two (after combining the data of both experiments).

Unsurprisingly, the lepton asymmetry in the CM frame is a more
sensitive probe than the lab frame lepton asymmetry.  However, we
emphasize that even the simple lab frame variable, which requires no
leptonic top reconstruction and is consequently free of many systematics,
can (1) help to establish the existence of an asymmetry inconsistent
with SM predictions, and (2) begin to distinguish between competing
models for the asymmetry.\footnote{Defining the visible
mass as the invariant mass of the lepton plus the 4 hard jets
identified as the visible $\ttbar $ decay products, enables efficient
isolation of high $\sqrt{s}$ events without reconstructing the
leptonic top.  Using a high {\em visible} mass bin $m_{vis} > 375$ GeV
yields a comparable enhancement to using a high total mass bin
$m_{t\bar t} > 450$.} 

The relationship between the lepton asymmetry and the parent top
asymmetry is a distinctive feature of the models: for the axigluon
models, which have similar top kinematics, the asymmetry is slightly
reduced due to kinematics for $G_A$, dramatically reduced for $G_L$,
and enhanced for $G_R$.  The $W'$, although similar to the $G_R$ in
yielding a higher proportion of right-handed tops, shows proportionally less of an
enhancement of the lepton asymmetry; this is because the $W'$ produces
tops which are more forward, where limited $y$ acceptance for leptons
causes events to fail acceptance cuts.

In \Tab{tab:tevsemilepeta} it is evident that the $W'$ model can be distinguished
from both the SM and the other reference models at $\gsim 3 \sigma$.
Our reference models $G_R $ and $W' $ predict similar central values 
for the lepton asymmetries but larger top asymmetries for the $W'$ model.
With the cuts in \cite{Aaltonen:2011kc}, the CM frame top asymmetries
differ by nearly $3\sigma $ for $G_R $ and $W' $.  The full anticipated
Tevatron data set allows a lepton asymmetry to be established at 
more than $3\sigma$ for all models except the $G_L$.
Meanwhile, the $G_L$ model has a very distinctive
signal of a lepton asymmetry near zero, which in the presence of a
significant top asymmetry is distinguishing.  
Moreover, the $G_A$ and $W'$ models can be discriminated at more than
$3 \sigma$.  One might worry that this distinction is artificial, as the 
$G_A$ model is chosen to underpredict the top asymmetry while the $W'$
model is not.  However, for comparison purposes we have also examined an axial axigluon parameter point
$G_A'$ with larger couplings (but the same mass and width) than the model 
displayed in Table II.  This model overpredicts the Tevatron $\ttbar$ cross-section
but has the same central values for the CM frame top asymmetries as the $W'$.  In
this modified $G_A'$ model, the lepton asymmetries are nearly $3 \sigma $ smaller
than in the $W'$ model.
Thus with the full anticipated data set
the Tevatron has the potential to discriminate between these explanations at
$\gsim 3 \sigma$ by exploiting the characteristic differences in the lepton asymmetries
relative to the asymmetry of the parent top.

\subsection{Dileptonic tops}

While dileptonic tops at the Tevatron are limited by both statistical
and kinematic reach, results from this channel are interesting
especially in combination with the semileptonic channel.  For
dileptonic tops, we impose selection cuts after
\cite{CDF-leptonic:2011}.  Specifically, we require two opposite sign
leptons, with $p_T >20$ GeV and $|\eta | <1.1$ for electrons, $|\eta |
<0.6 $ for muons; at least two jets satisfying $p_T >15$ GeV and
$|\eta | <2.5$; and large missing energy, $\met > 50~{\rm GeV}$.  In
addition a cut is placed on the scalar sum of the transverse energy of
the leptons, missing energy, and jets, $H_T \equiv \met +p_{T\ell_i} +
p_{Tj_i} >200 $ GeV.  After these cuts, the dominant backgrounds are
fakes, followed by Drell-Yan production of lepton pairs; signal makes up
$70\%$ of the events passing the cuts.

At the Tevatron, the most useful variables in
dileptonic tops are again (1) the polarization angle
$\cos\theta_\ell$ (see \Tab{tab:tevpol1}) and (2) the dileptonic charge asymmetry
of Eq.~(\ref{eq:ADeltaEll}).  
%
\begin{table}
\caption{New physics contributions to the dileptonic asymmetry. Results are shown
for reference models after imposing CDF dileptonic acceptance cuts as
in \cite{CDF-leptonic:2011}.  Statistical significances are based on the
number of signal events observed in \cite{CDF-leptonic:2011}. 
\label{tab:DilepADeltaEll}}
\begin{center}
\begin{tabular}{cl|c|c}
\hline \hline
       & mass  & asymmetry & stat.   \\
       & range &  (5.1 fb$^{-1}$)& sig. \\
\hline


$G_A$ & sel. cuts  & 8 \% & 1.2 \\
      & $m_{t\bar t}>450~{\rm GeV}$  & 14 \% & 1.4 \\
\hline

$G_L$ & sel. cuts   &  -4 \%  & 0.5  \\
      & $m_{t\bar t}>450~{\rm GeV}$  &  1 \%   & 0 \\

\hline 

$G_R$ & sel. cuts   & 15 \% & 2.4 \\
      & $m_{t\bar t}>450~{\rm GeV}$  & 20 \% & 2.1  \\
 
\hline

$W'$ & sel. cuts & 15 \%   & 2.3   \\
     & $m_{t\bar t}>450~{\rm GeV}$ & 24 \%  & 2.6   \\

\hline \hline
\end{tabular}
\end{center}

\end{table}
%
In \Tab{tab:DilepADeltaEll}, we show the dileptonic charge asymmetry
in our reference models, and quote statistical significances
based on the number of events observed in 5.1 $\mathrm{fb}^{-1}$.
While the smaller branching ratios into the dileptonic channel
limit the statistical reach compared to the semileptonic channel,
with the full data set a leptonic asymmetry can be established
at more than $3\sigma$ for the right-handed models, $G_R $ and $W' $.  
Other dileptonic variables,
such as those we will consider for the LHC in the next section,
are less sensitive at the Tevatron.

\section{LHC Studies}
\label{sec:lhc}
We now turn our attention to the LHC to see what light it can shed
upon any new physics effects contributing to the top asymmetry
measured at the Tevatron.

At first sight, measuring a charge asymmetry might seem impossible: the LHC, unlike the Tevatron,
collides identical particles (i.e. $pp$ instead of $p\bar p$) and it
is unclear how an asymmetry might be observed in this setup.  However,
while the initial state particles are symmetric at the LHC, the
quark and anti-quark parton distributions within them are not, with
valence quarks dominating over sea quarks at high $x$.  Thus, if the
$t\bar t$ system is left-moving then it is more likely to come from a
left-moving quark and a right-moving anti-quark than the other way
around, and vice-versa~\cite{Kuhn:1998kw,*Antunano:2007da,Wang:2010du,*Xiao:2011kp,*Cao:2011ew,*Bhattacherjee:2011nr,Hewett:2011wz}. 

We will make use of this correlation between the boost of the $\ttbar$
system and the partonic forward direction and adapt the observables
defined in the previous section to a $pp$ collider.  We will first
show that, if the Tevatron excess is due to contribution from new
physics, it can be confirmed at least at $3\sigma$ level with the
first 5 fb$^{-1}$ of LHC data.  At the same time, purely leptonic
observables with only minimal dependence on event reconstruction are
effective in probing the new physics.  We will also find that
measurements of the top quark polarization and $t\bar t$ spin
correlations can serve as powerful tools to distinguish different new
physics scenarios.

%
\begin{table}
\begin{center}
\caption{The cross section, in pb, for leptonically decaying di-top events at a 7 TeV LHC. \label{tab:7tev}} 
\vspace{10 mm}
\begin{ruledtabular} 
\begin{tabular}{c|cccccccc}
& $G_A$ & $G_L$ & $G_R$ & $W'$ &SM   \\ \hline
Selection cuts & 1.3 &	1.4 &	1.4 &	2.1 &	1.3  \\
 $m_{t\bar t} > 450 ~{\rm GeV}$  & 0.7 &	0.7 &	0.7 &	1.3 &	0.6\\
 $|y(t)+y(\bar t)| > 2$ & 0.2 &	0.2 &	 0.2 &	0.3 & 0.2 \\
\end{tabular}
\end{ruledtabular}
\end{center}

\end{table}
\begin{table}
\caption{The top forward-backward asymmetry ($\tafb$) at a 7 TeV LHC, and in parenthesis the $1\sigma$ statistical uncertainties, (i.e. $1/\sqrt{N}$)  
 assuming $5~{\rm fb}^{-1}$ of data.  \label{tab:tAFB}}
\begin{center}
\begin{ruledtabular} 
\begin{tabular}{c|ccccccccccccc}
 & $G_A ($\%$) $ & $G_L ($\%$) $ & $G_R ($\%$) $ & $W'(\%) $ & SM($\%$) \\ \hline
Selection cuts   &3 &	2 &	4 &	 14 &	 1 $(\pm1.2)$ \\
$m_{t\bar t} > 450 ~{\rm GeV}$   &5 &	3 &	 6 &	20 &	0 $(\pm1.7)$ \\
 $|y(t)+y(\bar t)| > 2$  & 8 & 5 &	12 &	36 &	1 $(\pm3.2)$\\
\end{tabular}
\end{ruledtabular}
\end{center}
\end{table}

We first describe the details of our simulation and our choices of
selection cuts.  As in the previous section, parton-level samples are
again generated using the full $2\rightarrow 6$ processes in
Madgraph~\cite{Alwall:2007st}.  However, here we will present results
more sensitive to top reconstruction and thus a more accurate
simulation is warranted.  Therefore, we further shower the
parton-level events in
Pythia~\cite{Sjostrand:2006za,*Sjostrand:2007gs}, cluster the visible
particles into $0.1\times0.1$ calorimeter cells between $-5<\eta<5$,
and form $R=0.7$ anti-$k_T$~\cite{Cacciari:2008gp} jets using
Fastjet~\cite{Cacciari:Fastjet,*Cacciari:2005hq}\footnote{For
  comparison, parton level results are tabulated in
  Appendix~\ref{sec:tables}.}.  The jets and leptons in our
reconstructed events are then passed through a realistic top
reconstruction algorithm (see App.~\ref{sec:reco}) which is used to
calculate our observables.

To select dileptonic tops, we employ a set of selection cuts adapted
from \Ref{Khachatryan:2010ez}.  These cuts require two hard
($p_{T\ell} > 20~{\rm GeV}$), central leptons ($|y_\ell| < 2.5$) with
an invariant mass outside the $Z$ window ($76~{\rm GeV} <
m_{\ell^+\ell^-} < 106~{\rm GeV}$), two hard, central jets ($p_{Tj} >
30~{\rm GeV}$, $|y_j| < 2.5$), and missing energy ($\met > 25~{\rm
  GeV}$).  These cuts reduce the backgrounds to di-top production to
the ${\cal O} (20\%)$ level, and so in what follows we will ignore
them.

To further isolate the effects of new physics we place a cut on the
reconstructed $t\bar t $ invariant mass
\be
\label{eq:mttcutlhc}
m_{t\bar t} > 450~{\rm GeV}.
\ee
At large $m_{t\bar t}$, the behavior of the observables considered
later in this section depends further on the details of the underlying
new physics model. For example, the axigluon- like models typically
have a broad resonance centered between $1 - 2$ TeV, and the resulting
observables in the regime of $m_{t\bar t}> 1 $ TeV are sensitive to
the precise location and width of the resonance. Such dependence can
provide valuable information.  Instead of pursuing this further, we
focus instead on the more universal nonresonant portion of the $\ttbar
$ spectrum, imposing an {\it upper bound} $m_{t \bar{t}} < 1.5 $
TeV. This represents a conservative approach.

Finally, since the process $gg\rightarrow t\bar t$ has a symmetric
initial state it cannot contribute to the asymmetry in $t\bar t$
production measured at the Tevatron, and thus constitutes a background
to the $q\bar q\rightarrow t\bar t$ process we wish to observe.  Now,
as the gluon PDFs fall more rapidly at large $x$ than quark PDFs, the
gluon-initiated contribution to top pair production tends to be more
central than the $q\bar q$-initiated contribution.  To suppress the
symmetric background from $gg\to \ttbar$, we employ a cut
\be
\label{eq:rap_cut}
|y_t+y_{\bar t}| > 2,
\ee
which restricts $t\bar t$ production to the relatively forward
region. The cut is chosen so that the contributions to $\ttbar$
production from $gg$ and $q \bar{q}$ are roughly comparable in this
regime. In addition, it also increases the correlation between the
boost of the $\ttbar$ system and the direction of the incoming quark.

The total $\ttbar$ production cross sections for the reference models,
including both the Standard Model and new physics contributions, after
imposing these cuts successively are shown in \Tab{tab:7tev}. For
comparison, the Standard Model production rates are also displayed.
Assuming 5 fb$^{-1}$ of data, we estimate the $1\sigma = 1/\sqrt{N}$
fractional statistical uncertainties (shown in parenthesis).

\subsection{Top forward backward asymmetry $\tafb$}
\label{sec:tafblhc}

One can generalize the forward-backward asymmetry $\tafb$ to a symmetric ($pp$) collider\footnote{Related observables have 
been considered before in~\cite{Kuhn:1998kw,*Antunano:2007da,Wang:2010du,*Xiao:2011kp,*Cao:2011ew,*Bhattacherjee:2011nr,Hewett:2011wz}} by  defining it as 
\be
\tafb = \frac{N(0<\hat{\theta}_t < \pi /2) - N( \pi /2 <\hat{\theta}_t < \pi )}{N(0<\hat{\theta}_t < \pi /2) + N( \pi /2 <\hat{\theta}_t < \pi )},
\ee
where $\hat{\theta}_t$ is the production angle of the top quark in the
$\ttbar$ center of mass frame with respect to the direction of the
boost of the $\ttbar$ system. Note that the $1 \sigma$ estimated in
\Tab{tab:tAFB} can also be used as a measure of the deviation of an
$\tafb$ measurement from the Standard Model prediction which is
approximately zero. Based on the procedure described in
Appendix~\ref{sec:reco}, we reconstruct the kinematics of the top
quark and calculate the $\tafb$ in various models. The results are
shown in \Tab{tab:tAFB}.  It can be seen from \Tab{tab:tAFB} that in
all but one of the conservative reference models, the new physics
contribution to the $\tafb$ can be identified at the LHC at the level
of $\gtrsim 3 \sigma$.  The only exception is the $G_L$ model, whose
contribution to $\tafb$ can only be detected at a level just below $2
\sigma$. This is mainly due to the bias introduced by our selection
cut. As discussed in the previous section, the direction of the lepton
is (anti)correlated with the (left)right polarized tops.  Therefore,
the boost of higher energy tops satisfying Eq.~\ref{eq:mttcutlhc} will
necessarily result in harder (softer) leptons. Combining with the
lepton $p_T$ cut, this effect leads to the relative enhancement of the
signal from the $G_R$ model relative to the $G_L$ model. We will
observe the same bias in other observables discussed later in this
section.

Note that although the rapidity cut in Eq.~\ref{eq:rap_cut} does help
to enhance the observed asymmetry, the asymmetry is already visible
after imposing the cut on $m_{t\bar{t}}$. Omitting the stringent cut
on $t \bar{t}$ rapidity will certainly enhance the signal statistics,
as shown in \Tab{tab:7tev}.  As we will see in the rest of this
section, the relative merit of including the rapidity cut depends on
the specific observable under discussion.  Therefore, we will base our
conclusions only on the most effective set of cuts.

From \Tab{tab:tAFB}, there are also observable difference between
various new physics scenarios. In particular, the $W'$ gives the
largest asymmetry. This is consistent with the pattern which is
already noticeable for $\tafb$ at the Tevatron, but it is much more
striking at the LHC.  This relative enhancement is due to the fact
that in the $W'$ model the new physics scale is much lower than in
axi-gluon models, so the increase in center-of-mass energy greatly
increases the fraction of time spent near the Rutherford singularity.
The analogous situation for an axigluon would be probing near the pole
mass, but to present a conservative model-independent estimate we have
removed this singularity (indeed, were we to have included and used a
low mass, our axigluon results could have shown similarly large
effects). We will also observe the same pattern in the other
observables discussed later in this section.

Within the axigluon models, we expect to observe a smaller
$\tafb$. Even in this case, we can gather early indications of whether
the underlying new physics model is of the $G_A$ type. In the
following, we will turn to additional observables to help both in
enhancing the signal significance, and distinguishing among different
axigluon models.




%
\subsection{(Nearly) Purely Leptonic Observables}

The first set of observables we will consider are those which involve
only leptons, and thus have only a minimal dependence on event
reconstruction\footnote{Cuts like \Eq{eq:mttcutlhc} require
reconstruction of the $t\bar t$ system, and thus can affect the
distribution of even purely leptonic observables by biasing which
events are selected.}.  We consider the generalization of the dilepton
charge asymmetry to a symmetric collider:
\be
\label{eq:lepaslhc}
{\cal A}^\ell_{FC}=\frac{N(|y_{\ell^+}| > |y_{\ell^-}| ) - N(|y_{\ell^-}| > |y_{\ell^+}| )  } 
                         {N(|y_{\ell^+}| > |y_{\ell^-}| ) + N(|y_{\ell^-}| > |y_{\ell^+}|)  },
\ee
measuring a {\em forward-central} leptonic charge asymmetry.
We present the measurement of this observable in \Tab{tab:afbllhc}.
\begin{table}
\caption{The forward-central lepton asymmetry ${\cal A}^\ell_{FC}$ at the 7 TeV LHC, and in parenthesis the $1\sigma$ statistical uncertainties, (i.e. $1/\sqrt{N}$)  
 assuming $5~{\rm fb}^{-1}$ of data.\label{tab:afbllhc}}
\begin{center}
\begin{ruledtabular} 
\begin{tabular}{c|ccccccccccccc}
& $G_A ($\%$) $ & $G_L ($\%$) $ & $G_R ($\%$) $ & $W'(\%) $ & SM($\%$)  \\ \hline
Selection cuts   &2 &	0 &	5&	13 &	0  $(\pm1.2)$\\
$m_{t\bar t} > 450 ~{\rm GeV}$   &4 &	2 &	 7 &	19 &	-1 $(\pm1.7)$ \\
 $|y(t)+y(\bar t)| > 2$  &7 &	2 &	14 & 35 &	1  $(\pm3.2)$\\
\end{tabular}
\end{ruledtabular}
\end{center}
\end{table}
As at the Tevatron, the leptonic charge asymmetry is a powerful probe
of the existence of a BSM asymmetry.  As can be seen from
\Tab{tab:afbllhc}, it allows $\gsim 3\sigma$ discrimination of all
models except the $G_L$ from the SM with order a few inverse femtobarn
of data, and much sooner in the case of the $W' $.  Moreover, the
$G_R$ model becomes distinguishable from the others. Thus the
dileptonic charge asymmetry can establish the existence of a BSM
asymmetry in $\ttbar$ events in typical axigluon or $t$-channel vector
boson models in the expected 7 TeV run.

To further strengthen the case for new physics and distinguish between 
competing explanations of an asymmetry, we consider several other leptonic variables.
The combination of these variables provides a diagnostic suite of measurements which,
taken together, can distinguish between different models for the top AFB.

One useful variable is the asymmetry in the azimuthal angle
between the two leptons, $\Delta\phi$, which is $\pi$ when the two
leptons are back to back and zero  when they are aligned in the
transverse plane.  In \Eq{eq:phiasymmlhc} we construct an asymmetry
with this variable
\be
\label{eq:phiasymmlhc}
{\cal A}^{\ell\ell}_ {\Delta\phi} =\frac{N(\cos\Delta\phi_{\ell\ell} > 0) - N(\cos\Delta\phi_{\ell\ell} < 0)  } 
                       {N(\cos\Delta\phi_{\ell\ell} > 0) + N(\cos\Delta\phi_{\ell\ell} < 0)  }.
\ee
measuring how often the two leptons are on opposite sides of the
transverse plane (contributing to ${\cal A}^{\ell\ell}_ {\Delta\phi} <
0$) vs. how often they are on the same side (${\cal A}^{\ell\ell}_
{\Delta\phi}>0$).  Unlike the leptonic asymmetry constructed in
\Eq{eq:lepaslhc}, there is a kinematic reason for ${\cal
  A}^{\ell\ell}_ {\Delta\phi}$ to be biased to negative values.
However, as one can see in \Tab{tab:delphilhc}, the difference between
the various ${\cal A}^{\ell\ell}_{\Delta\phi}$ provides a useful
discriminant. In particular, it helps in distinguishing the signal of
the $W'$ and $G_R$ models from the Standard Model.  We note that this
variable is not as sensitive as the others considered, and is not
useful at the Tevatron where new physics is probed at lower energies
and with smaller statistics.

\begin{table}
\caption{The azimuthal angle asymmetry ${\cal A}^{\ell}_{\Delta\phi}$ at a 7 TeV LHC, and in parenthesis the $1\sigma$ statistical uncertainties, (i.e. $1/\sqrt{N}$)  
 assuming $5~{\rm fb}^{-1}$ of data \label{tab:delphilhc}}
\begin{center}
\begin{ruledtabular} 
\begin{tabular}{c|ccccccccccccc}
& $G_A ($\%$) $ & $G_L ($\%$) $ & $G_R ($\%$) $ & $W'(\%) $& SM($\%$)   \\ \hline
Selection cuts  & -26 &	-26 &	-28 &	-43 &	-24  $(\pm1.2)$\\
$m_{t\bar t} > 450 ~{\rm GeV}$  & -47 &	-47 &	-50 &	-62 &	-45  $(\pm1.7)$ \\
 $|y(t)+y(\bar t)| > 2$ & -45 &	-44 &	-49 &	-56 &	-45  $(\pm3.2)$\\ 
\end{tabular}
\end{ruledtabular}
\end{center}
\end{table}

\subsection{Top Polarization}

As one might expect, fully reconstructed observables are even more
powerful than those which use only the leptons.  Next we explore
top polarization measurements, which again require reconstruction of the top rest
frame.
For simplicity we consider two choices of polarization axis: (1) the beam axis, which
we now define relative to the boost of the $\ttbar$ system,
\be
\hat{n}_{\rm beam}=
\begin{cases}
+\hat{z} &{\rm if \ }y_t+y_{\bar t}>0 \\ 
-\hat{z}& {\rm if \ }y_t+y_{\bar t}<0
\end{cases}
\ee
and (2) the helicity axis, again defined as the top direction of
motion in the $t\bar t$ center of mass frame.  The asymmetry in
${\cos\theta_\ell}$ which measures the net polarization does not
need to be redefined for the LHC, so we have once again
\be
\label{eq:thetaasymmlhc}
{\cal P}_ {n} =\frac{N(\cos\theta_{\ell, n} > 0) - N(\cos\theta_{\ell, n} < 0))  } {N(\cos\theta_{\ell, n} > 0) + N(\cos\theta_{\ell, n} < 0)) }.
\ee
Results are tabulated for the beam axis in \Tab{tab:beamlhc} and for the 
helicity axis in \Tab{tab:helicitylhc}.

\begin{table}
\caption{Net polarization ${\cal P}_ {\rm b} $ in the beam basis at a 7 TeV LHC, and in parenthesis the $1\sigma$ statistical uncertainties, (i.e. $1/\sqrt{N}$)  
 assuming $5~{\rm fb}^{-1}$ of data\label{tab:beamlhc}}
\begin{center}
\begin{ruledtabular} 
\begin{tabular}{cccccccccccccc}
& $G_A ($\%$) $ & $G_L ($\%$) $ & $G_R ($\%$) $ & $W'(\%) $& SM($\%$)   \\ \hline
Selection cuts  & 4 &	-1 &	5 &	 9 &	2  $(\pm1.2)$\\
$m_{t\bar t} > 450 ~{\rm GeV}$  & 1 &	-4 &	4 &	 11 &	 0  $(\pm1.7)$\\
 $|y(t)+y(\bar t)| > 2$ & 2 &	-5 &	7 &	15 &	1  $(\pm3.2)$\\
\end{tabular}
\end{ruledtabular}
\end{center}
\end{table}
\begin{table}
\caption{Net polarization ${\cal P}_ {\rm h} $ in the helicity basis at a 7 TeV LHC, and in parenthesis the $1\sigma$ statistical uncertainties, (i.e. $1/\sqrt{N}$)  
 assuming $5~{\rm fb}^{-1}$ of data \label{tab:helicitylhc}}
\begin{center}
\begin{ruledtabular} 
\begin{tabular}{c|ccccccccccccc}
& $G_A ($\%$) $ & $G_L ($\%$) $ & $G_R ($\%$) $ & $W'(\%) $& SM($\%$)   \\ \hline
Selection cuts  & 1 &	-1 &	 4 &	18 &	1 $(\pm1.2)$ \\
$m_{t\bar t} > 450 ~{\rm GeV}$  & 2 &	-2 &	6 &	26 &	0  $(\pm1.7)$\\
 $|y(t)+y(\bar t)| > 2$ & 0 &	-4 &	3 &	19 &	-2  $(\pm3.2)$\\ 
\end{tabular}
\end{ruledtabular}
\end{center}
\end{table}
Polarization measurements are particularly useful for distinguishing among the various
axigluon models, which differ from each other chiefly in the chiralities of their 
couplings to top quarks.  Polarization measurements also are important for
distinguishing the $G_L$ model from the SM. The bias towards right-handed polarizations
is an effect of selection cuts preferentially passing the harder leptons which arise
from right-handed tops.

\subsection{Top Spin Correlation}
Finally, we present results on $t\bar t$ spin correlations
\be
{\cal A}^\ell_ {c_1c_2} =\frac{N(c_1c_2> 0) - N(c_1c_2< 0))  } {N(c_1c_2 > 0) + N(c_1c_2< 0))  }
\ee
where $c_1=\cos\theta_{\ell_1, n} $ and $c_2=\cos\theta_{\ell_2, n} $.  As with the previous section, we 
present results using two polarization axes: the beam axis in \Tab{tab:beamcorr}
and 
the helicity axis in \Tab{tab:helicitycorr}.  

As one can see from the tables, ${\cal A}^\ell_ {c_1c_2}$ is not as
sensitive to new physics effects as the observables explored in the
previous subsections and thus will require more luminosity before
yielding meaningful information.  Nonetheless, as this observable
probes independent information it should still be measured as it can
help to further narrow down the set of explanatory models.

\begin{table}
\caption{Spin correlation ${\cal A}^\ell_ {c_1c_2}$ in the beam basis at a 7 TeV LHC, and in parenthesis the $1\sigma$ statistical uncertainties, (i.e. $1/\sqrt{N}$)  
 assuming $5~{\rm fb}^{-1}$ of data \label{tab:beamcorr}}
\begin{center}
\begin{ruledtabular} 
\begin{tabular}{c|ccccccccccccc}
& $G_A ($\%$) $ & $G_L ($\%$) $ & $G_R ($\%$) $ & $W'(\%) $& SM($\%$)   \\ \hline
Selection cuts  &-2 &	-2 &	-2 &	-9 &	-1  $(\pm1.2)$ \\
$m_{t\bar t} > 450 ~{\rm GeV}$  & -3 &	-3 &	-2 &	-11 & 0  $(\pm1.7)$ \\
 $|y(t)+y(\bar t)| > 2$ & -5 &	-4 &	-1 &	-11 &	-1  $(\pm3.2)$\\
\end{tabular}
\end{ruledtabular}
\end{center}
\end{table}
\begin{table}
\caption{Spin correlation ${\cal A}^\ell_ {c_1c_2}$ in the helicity basis at a 7 TeV LHC, and in parenthesis the $1\sigma$ statistical uncertainties, (i.e. $1/\sqrt{N}$)  
 assuming $5~{\rm fb}^{-1}$ of data\label{tab:helicitycorr}}
\begin{center}
\begin{ruledtabular} 
\begin{tabular}{c|ccccccccccccc}
& $G_A ($\%$) $ & $G_L ($\%$) $ & $G_R ($\%$) $ & $W'(\%) $& SM($\%$)   \\ \hline
Selection cuts  & -2 &	-3 &	-2 &	7 &	-4  $(\pm1.2)$\\
$m_{t\bar t} > 450 ~{\rm GeV}$  & 1 &	0 &	1 &	12 &	-2  $(\pm1.7)$\\
 $|y(t)+y(\bar t)| > 2$ & 3 &	 0 &	0 &	12 &	3  $(\pm3.2)$\\
\end{tabular}
\end{ruledtabular}
\end{center}
\end{table}

\section{Conclusions}
\label{sec:conclusions}
As more analyses are completed, the anomalous top forward backward asymmetry observed at the Tevatron appears increasingly robust.  The
observed asymmetry has manifested itself across many channels, at different experiments, and appears to rise strongly with increasing 
$m_{t\bar t}$, suggestive of new physics.  While a healthy dollop of skepticism is always
warranted in interpreting this sort of anomaly, especially as it may be susceptible to subtle QCD effects, the measurement presents at the very least one of the most
compelling collider anomaly seen in recent years, and may be our first hint of the BSM effects we may hope to observe at the LHC.

Here we have considered various  models built to explain the measured value of the top asymmetry.  While all models yield an enhanced $t\bar t$  asymmetry,  they yield distinct mixtures of left- and right-polarized tops.  As the polarization of a top influences the kinematics of its decay products, especially the charged lepton, distributions built from the leptons in dileptonic and semileptonic $\ttbar$ events are powerful probes of the physics underlying the anomalous top forward backward asymmetry. Leptonic charge asymmetries are particularly attractive for these purposes.  The potentially large central values of the lepton asymmetries make them sensitive probes of deviations from the SM, while the relationship between the lepton asymmetry and its parent $\ttbar$ asymmetry is a useful diagnostic of the chiral structure of the model.  Published results for lepton charge asymmetries would be very useful for understanding the properties of any new physics contributing to the $\ttbar $ cross-section.

Turning to the LHC, we found that this machine, despite being a $pp$ collider, has excellent prospects for shedding light on the top forward-backward asymmetry.
We defined an asymmetry variable at the LHC where the forward direction is chosen event by event, exploiting the distinct kinematics of sea and valence quarks, and utilized it with a novel set of cuts designed to enhance the contributions of new physics.  With our cuts and realistic reconstruction procedure, we are able to firmly establish the existence of a large BSM asymmetry within the first 5 fb$^{-1}$ of data at the 7 TeV LHC.  This result is based only on dileptonic tops, and the approach can easily be extended to the semileptonic channel, which promises a larger cross-section and better statistics.  While a full detector-level simulation will be necessary to understand the ultimate LHC sensitivity, this approach seems promising and warrants further study.  

We further observed that leptonic variables, namely lepton charge asymmetries, top polarizations, and $t\bar t$ spin correlations, are extremely valuable tools in distinguishing between the various new physics models which have been proposed to explain the anomalous AFB.  Using only the nonresonant portion of the $\ttbar $ spectrum, we saw that a $7~{\rm TeV}$ LHC is able to distinguish amongst various BSM models in as little as $\sim5~{\rm fb}^{-1}$ of data.

\acknowledgments{We would like to thank Claudio Campagnari, Qinghong Cao, Zhenyu Han, Dilani Kahawala, Matt Schwartz, Matt Strassler, Tim Tait, Paul Tipton, Yanjun Tu, Brock Tweedie, Itay Yavin, and Junjie Zhu for useful discussions, and particularly to thank Steve Giddings for collaborations in the early stage of this project. D.K., L.-T.W., and T.L. would like to thank Aspen Center for Physics for its hospitality. J.S. would like to think SLAC for its hospitality.  The work of T.L. was supported in part by the Department of Energy under contract DE-FG02-91ER40618.
D.K. is supported by a Simons postdoctoral fellowship and by 
an LHC-TI travel grant. J.S. was supported in part by DOE grant DE-FG02-92ER40704.  L.-T.W. is supported by the NSF under grant PHY-0756966 and the DOE Early Career Award under grant DE-SC0003930. }

\appendix
\section{Fully Leptonic Top Reconstruction}
\label{sec:reco}
Fully leptonic di-top events yield two neutrinos, and so the top four-vector is not immediately accessible from the collider data.  However, the eight unknowns comprising the two unmeasured neutrino four-vectors can be obtained if one requires:
\bi
\item $p_\nu^2=0$
\item $(p_\nu+p_{l})^2=m_W^2$
\item $(p_\nu+p_{l}+p_b)^2=m_t^2$
\item $(p_{\nu_1}+p_{\nu_2})_ {\{x,y\}} =(\met)_{\{x,y\}}$
\ei
Each item above yields two conditions which must be satisfied, so that in total the eight conditions - enough to solve\footnote{Many techniques are available to numerically solve the eight constraint conditions.  For our work we use the methods presented in \Ref{Sonnenschein:2006ud}.} the system up to discrete ambiguities.

In practice, a reconstructed event can yield zero, two, or four neutrino solutions.  If no neutrino solutions are found it could be because the top and/or $W$ produced in the event was sufficiently off-shell to make solving the constraint equations (which take as input the particle pole masses) impossible, or because the jet four-momenta were smeared too far away from their true values.  Events with two or four solutions can arise from the ambiguities in the solution to the constraint equations, and when there is an ambiguity in which $b$ to assign to which lepton in computing $m_t$.

To choose the correct solution we employ a technique developed in~\cite{Dalitz:1991wa} that assigns each neutrino solution a weight,
\be
w=f(x){\bar f}({x})p(E_l ^\ast | m_t)p(E_{\bar l} ^\ast | m_t)
\ee
where $E_{l} ^\ast$ is the energy of the lepton in the reconstructed top quark rest frame, 
\be
p(E_{l} ^\ast | m_t)=\frac{4E_l^\ast m_t(m_t^2-m_b^2-2E^\ast_lm_t)}{(m_t^2-m_b^2)^2-m_W^2(m_t^2+m_b^2)-2m_w^4}
\ee
measures the likelihood that a top will decay into a lepton of rest frame energy  $E_{l} ^\ast$, and $f$ ($\bar f$) are the relevant 
proton (anti-proton) PDFs\footnote{We use the CTEQ6.1 PDFs~\cite{Stump:2003yu} for our analysis.}.  In essence, this reconstruction method 
picks the neutrino/top solution which, all things being equal, seemed most probable.  For more details on the experimental implementation 
of this method we refer the reader to \Ref{Boline:2010zz}.  We also note that the method presented in \Ref{Bai:2008sk} might provide 
a complementary procedure.

\section{Models}
\label{sec:models}
To gauge the reach of the polarization and correlation techniques
presented herein we have adopted a set of benchmark models.  Here we
will introduce these models and delineate their parameters.

As was mentioned in \Sec{sec:intro}, BSM models designed to explain
the top forward-backward anomaly generally fall into two classes: $s$
and $t$ channel.  The good agreement between observation and SM
predictions for the top pair production cross-section makes it
challenging to reproduce an asymmetry as large as the observed effect
solely from BSM processes. If, instead, the asymmetry arises from the
interference of BSM graphs with SM QCD pair production, a sizable
asymmetry can be realized without comparably large positive-definite
contributions to the top pair production cross-section.  To obtain an
asymmetric contribution from interference effects of the necessary
size, the BSM graphs must have the same leading color and Lorentz
structures as the $q\bar q\to t\bar t$ contribution to SM top pair
production.  From this condition, we determine that BSM particles
exchanged in the $s$-channel must be spin 1 color octets, while
particles exchanged in the $t$-channel can be either vectors or
scalars, and can be in any color representation of QCD that appears in
the product of two fundamentals or a fundamental and an
anti-fundamental, namely $8, \bar 3, 6,$ or $1$.

For an $s$-channel particle to contribute to the asymmetry at tree
level, its couplings to both initial and final states must be parity
violating; otherwise any asymmetry will only be generated at loop
level, as in QCD.  We will call all such color vector octet particles
{\it axigluons}, even through the term more properly refers only to
such particles when they have a purely axial coupling.  At the
Tevatron, the axigluon is not produced on-shell, $m_A \gg \sqrt{s}$;
in this regime, the leading contribution of the axigluon to the top
asymmetry is determined by the mass of the axigluon and the product of
axial couplings to the first and third generations, $A_{FB}^t\propto
-g_A^1g_A^3$.  Note the sign: to yield a positive asymmetry, the
couplings to first and third generations must have opposite signs.
Once the mass and the product $g_A^1g_A^3$ are fixed, the relative
strength of the first and third generation axial couplings, as well as
the vector-like couplings, can be varied, with only subleading effects
on the asymmetry.  Here, in order to highlight the importance of
polarization, we choose three representative models which are all
consistent with the Tevatron data, yet have their vector-like
couplings adjusted to yield (1) a pure axial model; (2) a model which
only couples to left-handed tops; and (3) a model which only couples
to right-handed tops (see \Tab{tab:aximodels}).

\begin{table}
\caption{The parameters used in our reference axigluon models.  
We fix the axigluon width to be 1.4 TeV \label{tab:aximodels}}.
\begin{center}
\begin{ruledtabular} 
\begin{tabular}{c|c|c|c|c|c|c|c|c}
Model & $M$ [TeV] & $\Gamma$ [TeV]& $g_1^A$  &$g_1^V$  & $g_3^A$  & $g_3^V$ & $g^A$  & $g^V$  \\
\hline
${G}_A$ & $2.0$ & 1.40 & -2.3 & 0.0 & 3.35 & 0.0  &/ & /\\
\hline
${G}_L$ & $2.0$ & 1.40 & -2.3 & 0.0 & 3.35 & 3.35 &/ & / \\
\hline
${G}_R$ & $2.0$ & 1.40 & -2.3 & 0.0 & 3.35 & -3.35 &/ & / \\
\hline
$W'$ &$0.40$& 0.04 & / & / & / & / & $-0.90$ & $0.90$ \\
\end{tabular}
\end{ruledtabular}
\end{center}
\end{table}

The axigluon width depends upon many quantities which are unrelated to the reproduction 
of the forward-backward asymmetry (such as the coupling to second generation quarks, and the coupling to the $b$);
for simplicity we use a large constant width $\Gamma_G = 1.4~{\rm TeV}$ for all our reference models.
As the LHC can access the axi-gluon pole mass (which is around $2~{\rm TeV}$) 
a substantial model dependence can arise if the events from this region are analyzed as part of our signal.
We therefore take a conservative approach, explicitly removing this high invariant mass region from our analysis (see the discussion in \Sec{sec:tafblhc}).

In $t$-channel models, the asymmetry is generated by the kinematic
structure of the amplitude, from the forward singularity of the
$t$-channel propagator.  In order to yield a large asymmetry without
large corrections to the cross-section, the intermediate BSM state(s)
must have dominantly flavor-off-diagonal couplings.  Intermediate $t$
channel states can be either scalar or vector; for a given additional
contribution to the total $\ttbar$ cross-section, vectors give a
greater enhancement to the asymmetry than do scalars.  Intermediate
vectors must couple to $\bar q q$ and therefore must be octets or
singlets, while intermediate scalars can couple to either $q q$ or
$\bar q q$ and hence present more options.  Large flavor-changing
couplings are easier to accommodate if the coupling is only to
right-handed quarks; $t$-channel models thus generically predict
substantial top polarization.  If the exchanged particle is
self-conjugate, such as a $Z'$, then it will also mediate same-sign
top production, $q q\to t t$, the nondetection of which poses a
stringent constraint.

As an example of this class of model, we adopt a flavor-off-diagonal
$W'$ which couples $d_R$ to $t_R$,
\be
g_A=-0.9,\ g_V = 0.9, m_{W'}=0.4~{\rm TeV}, \Gamma = 0.04~{\rm TeV};
\ee
this parameter set is one of the benchmark models of
\cite{Gresham:2011pa}.

The $t\bar t$ cross-section and (Tevatron) $A_{FB}^ t$ which our
benchmark reference models predict is presented in
\Tab{tab:modelresults} along with a comparison to the SM.
Results are shown at leading order, and without acceptance cuts; after
incorporating realistic acceptance cuts, the apparently large
cross-section predicted by the $W'$ reference model is brought into
better accord with measurements~\cite{Gresham:2011pa}.
Note that the axigluon reference models, while consistent with cross-section
measurements, somewhat underpredict the observed forward-backward asymmetry.  
We have chosen these points to be conservative: models which realize a larger 
asymmetry are even easier to distinguish from the SM and from each other using 
the techniques presented here.

\begin{table}[t]
\caption{The LO $\ttbar$ production cross sections at the Tevatron 
and the LHC for the benchmark models, along with the CM frame top 
asymmetry.  \label{tab:modelresults}}
\begin{center}
\begin{ruledtabular} 
\begin{tabular}{c|c|c|c}
Model & $\sigma_{tt}^{\rm Tevatron}$[pb]  & $\sigma_{tt}^{\rm LHC}$[pb] & $A_{fb}^{\rm Tevatron}$\\
\hline
SM & 5.6 & 89 & $0\%$\\
\hline
${G}_A$ &5.8 &91 & $14\%$  \\
\hline
${G}_L$ & 6.1 &95 & $13\%$\\
\hline
${G}_R$ & 6.1& 95 & $13\%$\\
\hline
$W'$ & 7.3 & 123 & $24\%$\\
\end{tabular}
\end{ruledtabular}
\end{center}
\end{table}



\section{Tabulated Results for Leptonic Observables}
\label{sec:tables}
Here we collect tabulated predictions for several leptonic observables
for the SM and our reference models at both the Tevatron and the LHC.
We show results for multiple angular variables, integrated to
construct asymmetries.

For polarization variables we show the integrated asymmetry,
\be
\label{eq:thetaasymmlhc}
{\cal A}^\ell_ {\cos\theta} =\frac{N(\cos\theta > 0) - N(\cos\theta < 0))  } {N(\cos\theta > 0) + N(\cos\theta < 0))  }.
\ee
For azimuthal correlations between leptons, we show
\be
{\cal A}^\ell_ {\Delta\phi} =\frac{N(\cos\Delta\phi_{\ell\ell} > 0) - N(\cos\Delta\phi_{\ell\ell} < 0))  } {N(\cos\Delta\phi_{\ell\ell} > 0) + N(\cos\Delta\phi_{\ell\ell} < 0))  },
\ee
and for spin correlations we present the integrated variable
\be
{\cal A}^\ell_ {c_1c_2} =\frac{N(c_1c_2> 0) - N(c_1c_2< 0))  } {N(c_1c_2 > 0) + N(c_1c_2< 0))  }.
\ee

For the Tevatron we collect here expected net polarizations in the beam (Tab.~\ref{tab:tevpol2}) and off-diagonal bases (Tab.~\ref{tab:tevpol3}).
Results are displayed at parton level, after acceptance cuts. To gauge the statistical significance of the asymmetries, note that in
5.3 fb$^{-1}$ CDF observes $976.7 \pm 91.2$ signal events passing
their semileptonic cuts~\cite{Aaltonen:2011kc}.  As the statistical
uncertainty on these asymmetry variables is to a good approximation
simply given by $1/\sqrt{N}$, we then obtain a statistical error of
3\% on asymmetry variables with current data.  For dileptonic tops, in
5.1 fb$^{-1}$, CDF observes $237.1\pm 14.7$ signal events passing 
cuts~\cite{CDF-leptonic:2011}, for a statistical error of 6.5\% on
asymmetries.   Going to the high mass region will reduce the number of events
by a factor of $50\%$--$60\%$.
This procedure gives more conservative estimates of statistical significance than using
the significance derived from the cross sections after our (parton level) selection cuts.

We also present parton-level results for dileptonic top
variables at the 7 TeV LHC.  Tab.~\ref{tab:tAFB_pl} shows the resulting $t\bar t$ asymmetry
and Tab.~\ref{tab:afbllhc_pl} presents the related leptonic charge asymmetry.  We further give 
the parton-level polarization and spin-correlation results: Tab.~\ref{tab:delphilhc_pl} shows the 
distribution of ${\cal A}^{ll}_{\Delta\phi}$, Tab.~\ref{tab:tp_beam_pl} gives the top polarization 
using the beam-basis, while Tab.~\ref{tab:tp_hel_pl} presents this in the helicity basis, Tab.~\ref{tab:ttc_beam_pl}
shows the $t\bar t$ spin correlations in the beam basis, and finally Tab.~\ref{tab:ttc_hel_pl} shows these in the helicity basis.

\cleardoublepage

\begin{table}
\caption{Net polarization ${\cal P}_{\rm b}$ in the beam basis 
at the Tevatron.
\label{tab:tevpol2}}
\begin{center}
\begin{ruledtabular} 
\begin{tabular}{c|cc|cc}
       & \multicolumn{2}{c}{Semileptonic}   & \multicolumn{2}{c}{Dileptonic} \\
       & sel. cuts & $m_{\ttbar}>450$ GeV & sel. cuts & $m_{\ttbar}>450$ GeV \\
\hline
SM    &  -3 \% (3 \%) & -9 \% (5 \%) &  -8 \%  (6.5\%) & -14 \% (10 \%)\\
$G_A$ &  -5 \% & -10 \% &  -5 \% & -7 \%\\
$G_L$ &  -2 \% & -3 \% &  7 \% & 9 \%\\
$G_R$ &  -6 \% & -13 \% & -17 \% & -25 \%\\
$W'$  & -11 \% & -19  \% & -12 \% & -21 \% \\
\end{tabular}
\end{ruledtabular}
\end{center}
\end{table}

\begingroup
\begin{table}
\caption{Net polarization ${\cal P}_{\rm off-d}$ in the off-diagonal basis at the Tevatron. \label{tab:tevpol3}}
\begin{center}
\begin{ruledtabular} 
\begin{tabular}{c|cc|cc}
       & \multicolumn{2}{c}{Semileptonic}   & \multicolumn{2}{c}{Dileptonic} \\
       & sel. cuts & $m_{\ttbar}>450$ GeV & sel. cuts & $m_{\ttbar}>450$ GeV \\
\hline
SM    & -14 \% (3 \%) & -15 \% (5 \%) & -17 \% (6.5 \%) & -17 \% (10 \%) \\
$G_A$ &  -15 \% & -15 \% & -17 \% & -17 \% \\
$G_L$ &  -11 \% & -7 \% & -13 \% & -10 \% \\
$G_R$ &  -17 \% & -19 \% & -20 \% & -23 \% \\
$W'$  & -24 \% & -30 \% & -24 \% & -30 \% \\
\end{tabular}
\end{ruledtabular}
\end{center}
\end{table}
\endgroup



\begin{table}
\caption{The cross section, in pb, for leptonically decaying di-top events at a 7 TeV LHC, and in parenthesis the $1\sigma$ uncertainties 
on an asymmetry measurement centered about zero assuming $5~{\rm fb}^{-1}$ of data (i.e. $1/\sqrt{N}$).\label{tab:lhcxsec}} 
\begin{center}
\begin{ruledtabular} 
\begin{tabular}{c|ccccccccccccc}
& $G_A$ & $G_L$ & $G_R$ & $W'$ &SM   \\ \hline
Selection cuts & 2.1 &	2.2 &	2.2 &	 3.3 &	$2.1(\pm1.0\%)$ \\
 $m_{t\bar t} > 450 ~{\rm GeV}$  & 1.1 &	1.2 &	1.2 &	2.2 &	$1.1(\pm1.3\%)$ \\
 $|y(t)+y(\bar t)| > 2$ &0.3 &	0.3 & 	0.3 &	0.5 &	$0.2(\pm3.2\%)$ \\
\end{tabular}
\end{ruledtabular}
\end{center}
\end{table}


\begin{table}
\caption{The top forward-backward asymmetry  ${\cal A}^{t\bar t}$ at a 7 TeV LHC \label{tab:tAFB_pl}}
\begin{center}
\begin{ruledtabular} 
\begin{tabular}{c|ccccccccccccc}
& $G_A ($\%$) $ & $G_L ($\%$) $ & $G_R ($\%$) $ & $W'(\%) $ & SM($\%$) \\ \hline
Selection cuts   &2 &	2 &	3 &	15 &	0 \\
$m_{t\bar t} > 450 ~{\rm GeV}$   & 4 &	3 &	5 &	21 &	1 \\
 $|y(t)+y(\bar t)| > 2$  & 8 &7 &	11 &	38 &	2 \\
\end{tabular}
\end{ruledtabular}
\end{center}
\end{table}


\begin{table}
\caption{The forward-central lepton asymmetry ${\cal A}^\ell_{FC}$ at a 7 TeV LHC \label{tab:afbllhc_pl}}
\begin{center}
\begin{ruledtabular} 
\begin{tabular}{c|ccccccccccccc}
 & $G_A ($\%$) $ & $G_L ($\%$) $ & $G_R ($\%$) $ & $W'(\%) $ & SM($\%$) \\ \hline
Selection cuts   &2 &	0 &	5 &	14 &	0 \\
$m_{t\bar t} > 450 ~{\rm GeV}$   & 3 &	1 &	6 &	20 &	0 \\
 $|y(t)+y(\bar t)| > 2$  & 6 &	2 &	14 &	37 &	0 \\
\end{tabular}
\end{ruledtabular}
\end{center}
\end{table}


\begin{table}
\caption{The azimuthal angle asymmetry ${\cal A}^{ll}_{\Delta\phi}$ at a 7 TeV LHC\label{tab:delphilhc_pl}}
\begin{center}
\begin{ruledtabular} 
\begin{tabular}{c|ccccccccccccc}
 & $G_A ($\%$) $ & $G_L ($\%$) $ & $G_R ($\%$) $ & $W'(\%) $ & SM($\%$) \\ \hline
Selection cuts  &-23 &	-24 &	-26 &	-40 &	-21 \\
$m_{t\bar t} > 450 ~{\rm GeV}$  & -43 &	-44 &	-46 &	-58 &	-41 \\
 $|y(t)+y(\bar t)| > 2$ &-40 &	-45 &	-48 &	-54 &	-41 \\
\end{tabular}
\end{ruledtabular}
\end{center}
\end{table}

\begin{table}
\caption{Net polarization ${\cal P}_ {\rm b} $ in the beam basis at a 7 TeV LHC \label{tab:tp_beam_pl}}
\begin{center}
\begin{ruledtabular} 
\begin{tabular}{c|ccccccccccccc}
 & $G_A ($\%$) $ & $G_L ($\%$) $ & $G_R ($\%$) $ & $W'(\%) $ & SM($\%$) \\ \hline
Selection cuts  & 0 &	-2 &	3 &	6 &	0 \\
$m_{t\bar t} > 450 ~{\rm GeV}$  & 0 &	-2 &	3 &	9 &	-1 \\
 $|y(t)+y(\bar t)| > 2$ & 1 &	-8 &	9 &	13 &	-1  \\
\end{tabular}
\end{ruledtabular}
\end{center}
\end{table}


\begin{table}
\caption{Net polarization ${\cal P}_ {\rm h} $ in the helicity basis at a 7 TeV LHC\label{tab:tp_hel_pl}}
\begin{center}
\begin{ruledtabular} 
\begin{tabular}{c|ccccccccccccc}
 & $G_A ($\%$) $ & $G_L ($\%$) $ & $G_R ($\%$) $ & $W'(\%) $ & SM($\%$) \\ \hline
Selection cuts  & 5 &	3 &	9 &	21 &	5 \\
$m_{t\bar t} > 450 ~{\rm GeV}$  & 6 & 2 &	11 &	28 &	6 \\
 $|y(t)+y(\bar t)| > 2$ & 8 &	 4 &	11 &	25 &	5 \\
\end{tabular}
\end{ruledtabular}
\end{center}
\end{table}


\begin{table}[t]
\caption{Spin correlation asymmetry ${\cal A}^\ell_ {c_1c_2}$ in the beam basis at a 7 TeV LHC \label{tab:ttc_beam_pl}}
\begin{center}
\begin{ruledtabular} 
\begin{tabular}{c|ccccccccccccc}
 & $G_A ($\%$) $ & $G_L ($\%$) $ & $G_R ($\%$) $ & $W'(\%) $ & SM($\%$) \\ \hline
Selection cuts  & -2 &	-2 &	-3 &	-6 &	-3 \\
$m_{t\bar t} > 450 ~{\rm GeV}$  & -5 &	-5 &	-5 &	-9 &	-5 \\
 $|y(t)+y(\bar t)| > 2$ & -7 &	-8 &	-10 &	-10 &	-8 \\
\end{tabular}
\end{ruledtabular}
\end{center}
\end{table}


\begin{table}[t]
\caption{Spin correlation asymmetry ${\cal A}^\ell_ {c_1c_2}$ in the helicity basis at a 7 TeV LHC  \label{tab:ttc_hel_pl}}
\begin{center}
\begin{ruledtabular} 
\begin{tabular}{c|ccccccccccccc}
 & $G_A ($\%$) $ & $G_L ($\%$) $ & $G_R ($\%$) $ & $W'(\%) $ & SM($\%$) \\ \hline
Selection cuts  & -5 &	-4 &	-4 &	4 &	-6 \\
$m_{t\bar t} > 450 ~{\rm GeV}$  & -2 & -1 &	0 &	10 &	-3 \\
 $|y(t)+y(\bar t)| > 2$ & 0 &	 1 &	3 &	10 &	0 \\
\end{tabular}
\end{ruledtabular}
\end{center}
\end{table}

\cleardoublepage

\bibliography{tafb}
\bibliographystyle{apsrev4-1}
\end{document}
\end{document}